\documentclass[titlepage,onecolumn,12pt,nofootinbib,prd,floatfix,preprintnumbers,amsmath,amssymb,groupedaddress]{revtex4}
\usepackage{epsfig}  
\usepackage{array}  
\usepackage{axodraw}  
\newcommand{\TeV}{\,\mathrm{TeV}}  
\newcommand{\GeV}{\,\mathrm{GeV}}

\newcommand{\eV}{\,\mathrm{eV}}  
\begin{document}   
\title{Neutrinos in a Sterile Throat} 
\date{\today}
\author{Ben Gripaios}  
\email{b.gripaios1@physics.ox.ac.uk}  
\affiliation{Rudolf Peierls Centre for Theoretical Physics, University of Oxford,
1 Keble Rd., Oxford OX1 3NP, UK}
\affiliation{Merton College, Oxford OX1 4JD, UK}
\begin{abstract}
We consider field-theoretic models of a warped extra dimension with multiple throats, in which fermions that
 are singlets of the Standard Model gauge group propagate in a separate throat from the Standard Model fields,
 which we call the sterile throat.
  The singlets mix with Standard Model fields via interactions localized on the UV brane that connects the two throats.
This leads to three, light, \mbox{mostly-active}, Majorana neutrinos via a higher-dimensional see-saw mechanism, 
together with Kaluza-Klein towers of mostly-sterile neutrinos, whose scale is set by the warp factor in the sterile throat
and can be very low if the throat is deep.
We suggest that a model of this kind may explain all the neutrino data, reconciling the LSND result with astrophysical constraints. 
\end{abstract}   
\pacs{}
\keywords{Extra-dimensional models, neutrino masses}
\preprint{OUTP-0619P}
\maketitle 
\section{\label{intro}Introduction}
The observation of neutrino masses and mixing has provided us with the first window
onto physics beyond the Standard Model (SM). At least inasmuch as the solar and atmospheric neutrino
experiments are concerned, the data are well described by the SM
 supplemented with two or more fermions transforming as singlets under the SM
gauge group \cite{Gonzalez-Garcia:2002dz,Mohapatra:2006gs}. These sterile neutrinos, $\nu_R$, interact at the renormalizable level
with the three active neutrinos, $\nu_L$, of the Standard Model and the Higgs, $H$, via a Yukawa interaction 
(which becomes
a Dirac mass term for the neutrinos once the Higgs gets a vacuum expectation value) and with themselves
via a Majorana mass term. Schematically, the SM Lagrangian is supplemented with the terms
\begin{gather}
\delta \mathcal{L} \supset \lambda H \nu_L \nu_R + m \nu_R \nu_R + \mathrm{H.\ c.},
\end{gather}
together with the usual kinetic terms for $\nu_R$. 
A good fit to the solar and atmospheric neutrino data is obtained either by taking the Majorana masses $m$
to be zero (in which case the Yukawa couplings $\lambda$ must also be very small and 
the mass eigenstates are Dirac fermions) or by choosing $m$ to be large (around $10^{14-15}$ GeV, 
in which case the $\lambda$ are of order unity and there are three, light, mostly-active mass eigenstates, with the rest very heavy).
 In the latter, see-saw model, the presence of SM-gauge-singlet fermions at a high scale explains why some 
 neutrino masses are so light.

There is, however, one short-baseline experiment, LSND, that proves to be more problematic.
The LSND experiment \cite{Aguilar:2001ty} observes $\overline{\nu}_e$ appearance in a $\overline{\nu}_{\mu}$ beam at more than $3\sigma$ above background, which
can be fit by a two-neutrino oscillation with a mass-squared difference of around an $\mathrm{eV}^2$. 
If the LSND result and the oscillation hypothesis are correct
(the MiniBoone experiment, which will independently probe the LSND region, is expected to announce its results imminently)
 then we  have evidence
for three independent mass-squared differences in the low-energy neutrino spectrum, {\em viz.\ }$\Delta m^2_{sol} \sim 10^{-5} \mathrm{eV}^2$,
$\Delta m^2_{atm} \sim 10^{-3} \mathrm{eV}^2$ and $\Delta m^2_{LSND} \sim \mathrm{eV}^2$. 
Na\"{\i}vely, these are easily accommodated in the minimal extension of the SM with sterile neutrinos and non-vanishing Majorana masses
discussed above, by choosing some of the mostly-sterile mass eigenstates to have mass of order an eV.
The oscillations relevant for LSND then occur via one or more intermediate, mostly-sterile states. 
It turns out that at least two such states are required for an adequate fit to the data 
\cite{Maltoni:2002xd,Strumia:2002fw,Sorel:2003hf,Schwetz:2003pv,Maltoni:2004ei,deGouvea:2005er}, 
because of the constraints coming from a number of other
short-baseline experiments that do not see any evidence for oscillations.

The objections to this explanation of LSND are both theoretical and observational. 
From a theoretical viewpoint, the explanation of LSND using the minimal extension
of the SM is unsatisfactory, in that it introduces a light (eV) mass scale in an {\em ad hoc} fashion.
We stress, though, that this objection is purely aesthetic: all values of the fermion masses are technically natural,
and so assigning arbitrary values to them is no worse than assigning arbitrary values to the Yukawa couplings of quarks
and charged leptons in the Standard Model. Still, an explanation of these values must, ultimately, be forthcoming.

The observational objection is that extra, light neutrino states are in conflict with standard Big Bang cosmology.
This objection is more serious, and has become acute 
following the recent measurement of the cosmic microwave background radiation by WMAP \cite{Spergel:2006hy}.
Extra, light neutrino states have two important effects in the early Universe.
Firstly, they affect the predicted primordial abundance of light elements (in particular $^4\mathrm{He}$) 
that result from
Big Bang nucleosynthesis (BBN), because BBN is sensitive to the energy density stored in relativistic species,
and because BBN depends on reactions involving neutrinos, such as $\overline{\nu}_e p \rightarrow e^+ n$.
Using the WMAP data, which provides a 
precise measurement of the baryon-to-photon ratio, the effective number of
thermalized, relativistic neutrinos at BBN is given by \cite{Pierce:2003uh,Cirelli:2006kt,DiBari:2001ua,DiBari:2003fg} 
$$N_{\nu}^{\mathrm{eff}}<3.4, \;\ \;\; 95\% \;\;\mathrm{(two-sided)\;\; C.\ L.}$$
It thus appears difficult to accommodate even one extra neutrino, let alone two.
Secondly, neutrinos affect structure formation, since they tend to decouple
and free-stream well before structure begins to form. The spectrum of initial
density perturbations is, therefore, cut-off on the smallest scales \cite{Kolb:1990vq}.
The WMAP data, together with observations of large-scale structure, then provide a constraint
on the energy density stored in neutrinos (more precisely, on $\Omega_{\nu} h^2$) and {\em ergo} on the sum of neutrino
masses. The constraint is model-dependent; for a model with four or five neutrinos, of which three are degenerate, the 95 \% confidence level bounds
 are \cite{Hannestad:2003xv} 
 \begin{align} \label{summ}
 \Sigma m_{\nu} < 1.4 \mathrm{eV}, \;\; N_{\nu} = 3 +1,\nonumber \\
 \Sigma m_{\nu} < 2.1 \mathrm{eV}, \;\; N_{\nu} = 4 +1.
 \end{align}
 Again, it would appear that multiple neutrino states at an eV or above are only marginally possible.
It seems then, that a coherent explanation of all the neutrino data is hard to find.\footnote{Other approaches to explain the LSND result include $CPT$-violation \cite{Murayama:2000hm}, violation of Lorentz invariance \cite{Kostelecky:2003cr},
 sterile neutrino decay \cite{Palomares-Ruiz:2005vf}, and decoherence \cite{Barenboim:2004wu}.}
 This state of affairs will become particularly embarrassing if MiniBoone confirms the LSND result.

In this work, we wish to suggest a radical approach to sterile neutrinos. Our starting point is the observation that both
objections to extra, light sterile neutrino states may be surmountable
in a model where the extra states are composite degrees of freedom. 
As far as the theoretical objection is concerned, we can explain the existence of an eV scale in a natural way by supposing that
the required sterile neutrinos are not fundamental fields of an effective field theory, valid up to some high energy cut-off,
but rather are composite bound states of some theory whose couplings grow slowly in the infra-red, and 
become strong around an eV. This is entirely analogous to what happens in the strong interactions,
where the QCD gauge coupling increases logarithmically in the IR, such that the strong-coupling scale is around a GeV
and the resulting composite bound states are the hadrons. In this way, we find a natural
explanation for the existence of a scale that is very light compared to other scales in Nature:
it arises because the couplings run so slowly.\footnote{The idea of compositeness to explain neutrino mass scales was used before in \cite{Arkani-Hamed:1998pf}.}

Moreover, the supposition that sterile neutrinos are composite means that
the discussion of the effects of the neutrinos on physics in the early Universe is changed.
Indeed, at early times (in fact, at all energy scales much greater than the compositeness scale),
the weakly-coupled degrees of freedom of the theory do not consist of the neutrinos {\em per se}, but rather the degrees of freedom
appropriate to the ultra-violet (UV) completion of the theory, whatever that may be.

In the absence of a calculable UV completion, this observation is of limited use.
Remarkably, such a completion can be found, by recourse to the celebrated
AdS/CFT correspondence \cite{Maldacena:1997re}. In the correspondence, or rather a regularized version thereof
 \cite{Arkani-Hamed:2000ds,Rattazzi:2000hs,Perez-Victoria:2001pa},
 the UV completion consists of a 4-d conformal field theory of large rank, coupled linearly to dynamical source fields. The theory 
becomes strongly-coupled in the infra-red (IR), triggering spontaneous breaking of the conformal symmetry, and the appearance of a discrete spectrum of composite bound states.
According to the correspondence, this theory is holographically dual
to a weakly-coupled (and therefore calculable) 5-d theory formulated on a slice of an $AdS_5$ geometry, that is terminated by branes located at the points 
$z=z_0$ and $z=z_1$ in the fifth dimension. Going back to the 4-d theory, the extra-dimensional co-ordinate $z$ is interpreted as the renormalization group scale, and the branes translate to 
cutting off the CFT in the UV at a scale $1/z_0$, and  
to strong-coupling occurring in the IR at a scale $1/z_1$. To an observer living in four dimensions, the 5-d fields manifest themselves as Kaluza-Klein (KK)
towers of 4-d mass eigenstates, the mass spectrum beginning at around $1/z_1$. These KK states are dual
to the composite objects that arise when the conformal symmetry is spontaneously-broken in the IR.

Theories of this type have already been extensively invoked in the literature to explain the phenomenology of the SM,
in particular the naturalness of its electroweak sector 
\cite{Randall:1999vf,Randall:1999ee,Contino:2003ve,Agashe:2004rs}. The SM fields are assumed to propagate in a slice of $AdS_5$ (which we call the SM throat). By arranging the Higgs field in such a way
that its lightest mode is localized towards the $z_1$ brane, where $1/z_1 \sim \mathrm{TeV}$, one obtains a solution to
the gauge hierarchy problem: in the 4-d dual, the Higgs is wholly or predominantly a composite object of the CFT, so its mass parameter 
is insensitive to radiative corrections occurring on distance scales much shorter than an inverse TeV.

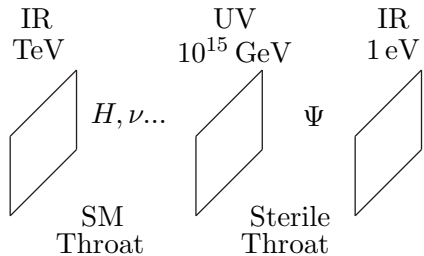
\begin{figure}[t]  
\begin{center}  
\begin{picture}(120,60)(0,0)  
\Line(0,10)(0,40)  
\Line(0,10)(25,35) 
\Line(0,40)(25,65)
\Line(25,35)(25,65)  

\Line(70,10)(70,40)  
\Line(70,10)(95,35) 
\Line(70,40)(95,65)
\Line(95,35)(95,65)  

\Line(130,10)(130,40)  
\Line(130,10)(155,35) 
\Line(130,40)(155,65)
\Line(155,35)(155,65)  
\Text(10,73)[]{$\TeV$}  
\Text(45,48)[]{$H,\nu...$}
\Text(35,10)[]{SM} 
\Text(34,0)[]{Throat}
\Text(10,85)[]{IR}
\Text(85,73)[]{$10^{15}\GeV$}
\Text(115,48)[]{$\Psi$}
\Text(85,85)[]{UV}
\Text(106,10)[]{Sterile}
\Text(104,0)[]{Throat}
\Text(145,73)[]{$1\eV$}
\Text(145,85)[]{IR}
\end{picture}  
\end{center}  
\caption{The two-throat set-up, with SM fields ($H,\nu,\dots$) and sterile neutrinos ($\Psi$) living in 
separate throats and interacting at a common UV brane.}  
\label{throat}  
\end{figure}

Coming back to the issue of sterile neutrinos, we see that if we wish to explain the existence of an eV scale in this way,
the sterile neutrinos must live in a slice of $AdS_5$ with $1/z_1 \sim \mathrm{eV}$. We immediately see that
the SM fields cannot propagate in the bulk of the same throat, because there would then be KK excitations of all SM fields beginning at around an eV,
in clear contradiction with experiment. The sterile 
neutrinos must therefore live in a throat of their own, which we call the sterile throat,
and interact with fields in the SM throat only through interactions localized on the UV brane that 
connects the two.\footnote{The SM fields need not live in a throat at all:
they could of course be simply localized on the UV brane of the deep throat, with the gauge hierarchy problem solved in some other way.}
This set-up is illustrated in Figure \ref{throat}.
%
Such a scenario of field theories in multiple $AdS_5$ throats has been recently explored in \cite{Cacciapaglia:2006tg,Brummer:2005sh,Hebecker:2006bn}, in which multiple slices of $AdS_5$ space are connected via a common UV brane. It has a natural dual 
explanation in terms of separate CFTs that become strongly-coupled 
at their respective IR brane scales and interact with each 
other indirectly via external fields living on the common UV brane. The set-up also provides a toy model of recent string theory compactifications \cite{Douglas:2006es}
on Calabi-Yau manifolds with fluxes.

In what follows, we perform a detailed analysis of the set-up illustrated in Figure \ref{throat}.
We begin, in Section \ref{1BN}, by developing the theory of a single bulk sterile neutrino propagating in a slice of AdS.
A gauge-singlet fermion has both Dirac and Majorana mass terms in the warped bulk. The general case where neither of these
vanishes has received little attention in the literature thus 
far (a brief numerical analysis with sterile neutrinos in the SM throat was performed in \cite{Huber:2003sf}).\footnote{For related references, see \cite{Dienes:1998sb,Arkani-Hamed:1998vp,Dvali:1999cn,Grossman:1999ra,Lukas:2000rg,Hebecker:2002re,Hebecker:2002xw,Kim:2002im,Gherghetta:2003he,Haba:2006dz}.}
There is perhaps good reason for this: we show that the decoupled, second-order, differential equations
for the extra-dimensional profiles of the sterile neutrino KK modes are not of confluent hypergeometric form.
They are, however, Fuchsian at the origin, and we define the solutions by the two independent series solutions there.
Unlike in the case of pure Dirac masses \cite{Contino:2004vy}, we show that massless or ultralight modes, 
with tunable localization properties, do not occur for natural choices of the boundary conditions
on the UV and IR branes. 
Instead, one typically obtains a KK spectrum of Majorana fermions with masses beginning at 
around $1/z_1$.\footnote{The co-ordinate $z$ will always refer to the sterile throat in what follows.} 

We are most interested in the particular scenario where the bulk sterile neutrino has an interaction with the Standard Model
fields via a Yukawa-type interaction, localized on the UV brane, connecting the active neutrinos, the Higgs and the bulk sterile 
neutrino.\footnote{We have in mind more sophisticated realizations of the Randall-Sundrum scenario,
in which the Higgs is not $\delta$-function localized on the IR brane, but instead propagates in the bulk of the SM throat \cite{Contino:2003ve,Agashe:2004rs,Davoudiasl:2005uu}.
These offer a number of advantages relative to the original Randall-Sundrum example.}
We calculate, in \ref{bc}, the boundary conditions on the UV brane that follow from this interaction, and show that they
lead to the presence of an ultralight mass eigenstate in the KK spectrum, with mass of order $\mathrm{TeV}^2 z_0$, which is mostly made up of active neutrinos.
Plausible values for the light neutrino masses result for $1/z_0$ in the region of $10^{14-15}$GeV. 
The magnitude of the ultralight mass is just what one would obtain in the usual see-saw mechanism with fundamental sterile neutrinos with mass of $O(1/z_0)$,
but results from the mixing of the active neutrino with the entire KK tower of sterile neutrinos, and the specific expression for the mass reflects this.
The spectrum of higher KK states begins at around $1/z_1$ and is made up of mostly sterile states. 
At large KK number, the states are separated in mass by $\pi /z_1$.

In Section \ref{>1BN}, we generalize to more than one bulk sterile neutrino. 
In this case, the analysis is further complicated by the observation that no basis exists, in general, in which
the bulk Dirac and Majorana mass terms 
are simultaneously diagonal. This prevents us from giving a simple expression for the masses
of the ultralight modes.
(We do, however, present a simple expression for the masses of the ultralight modes in the special case where the mass
terms can be diagonalized  simultaneously.) There are three ultralight modes (one for each active neutrino of the SM), predominantly composed of active neutrinos,
with order-one mixing between the SM flavour eigenstates. 
As a result of the mixing, neutrino oscillations occur between the active neutrino flavour eigenstates.
The solar and atmospheric data can be explained by oscillations between the mostly-active ultralight modes.
We further conjecture that the LSND result can be explained by 
oscillations between the mostly-active ultralight states and the mostly-sterile higher KK states.
A detailed investigation of this requires knowledge of the spectrum of higher KK modes and hence a numerical analysis,
which is beyond the scope of this work.

In Section \ref{highp}, we address the high-energy behaviour of the theory, with the early Universe in mind, reverting to
the case of a single bulk sterile neutrino for simplicity. 
The theory is well-defined and calculable (via its 5-d gravity dual) at energies up to and beyond the scale $1/z_0$
(up to the fundamental 5d gravity scale in fact, which we take to be $M_5 \sim 10^{16-17} \mathrm{GeV}$, so as to
obtain the correct 4d gravity scale, $M_4^2 \sim M_5^3 /k$).

The crucial consideration, as far as the early Universe is concerned, is whether the extra, mostly-sterile states composing the KK tower
can come into equilibrium with the primordial plasma of SM excitations before the epochs of BBN and structure formation.
If they do (either wholly or partially), then their contribution must be included in the standard framework. 
We envisage two processes by which equilibriation might occur. One is by interactions in the CFT itself, and the other is by SM interactions.
In either case, the actual communication between the SM and sterile CFT sectors occurs not at the interaction vertex, but via the mixing in the propagators, or equivalently in the mass eigenstates.
On the one hand, we argue that the extra states cannot come into equilibrium by CFT interaction processes.
On the 5d side, the explanation for this is that the KK states 
are localized in the IR, and can only interact with the SM fields localized on the UV brane through the 5d UV-to-IR brane propagator
for some field that propagates in the bulk of the sterile throat. Now, 
at large Euclidean four-momenta, these propagators fall off exponentially above the KK scale, as $e^{-pz_1}$. 
On the 4d side, the explanation is that, an Euclidean momenta far above $1/z_1$, conformal symmetry is restored, and the composite states
are transparent to probes on short momentum scales \cite{Gherghetta:2003wm}.
Thus, even though the number of KK states that are above threshold for some process and the available phase space
grow as some powers of $p$, these enhancement factors will always be overwhelmed by the exponential suppression at large
enough $p$, and interactions between SM fields and the extra neutrino states are negligible at energies much above an eV.

On the other hand, the extra states could equilibriate via SM interactions, just as standard 4-d sterile neutrinos can \cite{Barbieri:1989ti}.
These processes are not suppressed at high energies. As for laboratory oscillations, an understanding
of the size of this effect will require numerical analysis, which is not carried out here. We hope to perform
an analysis of this, and other phenomenological effects, in future work.

Finally, in the Discussion, we address the feasibility of performing a global fit to the data and the implications for future
oscillation experiments, neutrinoless double-beta decay experiments, and astrophysics.
\section{\label{1BN}One Bulk Sterile Neutrino}
\subsection{Formalism}
We realize a single throat as a slice of an $AdS_5$ geometry in Poincare 
co-ordinates\footnote{We use lower-case Greek letters for 4-d spacetime indices, 
lowercase Latin for 3-d space indices and uppercase Latin for 5-d spacetime indices.} $X^M = (x^\mu, z)$, where 
$z \; \epsilon \;[z_0 = 1/k, z_0 = e^{\pi k R}/k]$. The metric (signature mostly-plus) is 
\begin{gather}
ds^2 = \frac{1}{(kz)^2}(\eta_{\mu \nu} dx^{\mu} dx^{\nu} + dz^2),
\end{gather}
with {\em funfbein} $e^M_A = kz \delta_A^M$.
The metric is a solution of General Relativity in a 5-d space with negative cosmological constant $\Lambda_5$, 
terminated by 3-branes of positive and negative tension located at $z_0$ and $z_1$, respectively.
The AdS curvature scale is given by
\begin{gather} \label{bulkcc}
k^2 = - \frac{\Lambda_5}{12 M_5^3},
\end{gather}
where $M_5$ is the 5-d gravity scale. The brane tensions, $\Lambda_4$, are given by
\begin{gather} \label{tension}
k =  \frac{|\Lambda_4|}{6 M_5^3}.
\end{gather}

For a multi-throat geometry \cite{Cacciapaglia:2006tg}, we splice together slices of AdS (with fifth-dimensional co-ordinates $z,y,\dots$)
by identifying the point $z_0 = 1/k$ with $y_0 =1/l$ and so on, corresponding to a common UV brane. The
curvature scales $k,l,\dots$ are set by the values that the cosmological constant takes in the bulk of each throat,
as per (\ref{bulkcc}). Applying the Israel junction condition on the UV brane leads to the relation
\begin{gather} \label{t}
k+l+\dots =  \frac{|\Lambda_4|}{6 M_5^3}.
\end{gather}

The five-dimensional Dirac algebra, $\{\gamma^A,\gamma^B\}=2 \eta^{AB}$, is represented by
\begin{align}
\gamma^{\mu} &= 
\begin{pmatrix}
0 & i\sigma^{\mu} \\ i\overline{\sigma}^{\mu} & 0
\end{pmatrix}, \nonumber \\
\gamma^5 & =
\begin{pmatrix}
-1 & \phantom{-}0 \\ \phantom{-}0 & \phantom{-}1
\end{pmatrix},
\end{align}
where $\sigma^{\mu} = (1, \sigma^i)$, $\overline{\sigma}^{\mu} = (1, - \sigma^i)$, and $\sigma^i$ are the usual Pauli matrices.
The 5-d charge conjugation matrix 
\begin{gather}
C_5 = \begin{pmatrix}
-\sigma^{2} & 0 \\ 0 & - \sigma^{2}
\end{pmatrix},
\end{gather}
satisfies the relation
\begin{gather}
C_5 \gamma^A {C_5}^{-1} = + (\gamma^A)^t,
\end{gather}
where $t$ denotes the transpose.
\subsection{The Bulk Action and Equations of Motion}
The smallest irreducible representation of the 5-d Dirac algebra is carried by a four-component Dirac spinor $\Psi$.
The most general Hermitian and 5-d Lorentz-invariant action for $\Psi$ is then
\begin{gather}
S = - \int d^4 x dz\; \sqrt{-g} \frac{i}{2}\Big( \overline{\Psi}\Gamma^M D_M \Psi -  ck \overline{\Psi}\Psi
- dk \Psi^t C_5 \Psi \Big) + \mathrm{H.\ c.},
\end{gather}
where $\Gamma^M = e^M_A \gamma^A$ are curved-space gamma matrices and $D_M = \partial_M + \omega_M$ includes the spin connection.
The parameters $c$ and $d$ are of order unity in a natural theory and we may take $c$ to be real, 
without loss of generality. In terms of the re-scaled Weyl spinors, $\psi_{\alpha} $and $ \overline{\chi}^{\dot{\alpha}}$, defined such that 
\begin{gather}
\Psi = (kz)^2 \begin{pmatrix} \psi_{\alpha} \\ \overline{\chi}^{\dot{\alpha}}  \end{pmatrix},
\end{gather} 
the action is
\begin{multline} \label{actpsi}
S  = - \int d^4 x dz \; \Big( 
-i\chi \sigma \cdot \partial \overline{\chi}
-i\overline{\psi}\overline{\sigma}\cdot \partial \psi
+\frac{1}{2}\left( \chi \partial_z \psi + \partial_z \overline{\psi} \overline{\chi} - \overline{\psi}\partial_z \overline{\chi}-\partial_z \chi \psi \right)\\
+\frac{c}{z}(\chi \psi + \overline{\psi} \overline{\chi}) 
+\frac{d}{2z}(\psi \psi - \overline{\chi} \overline{\chi}) +\frac{\overline{d}}{2z}(\overline{\psi}\overline{\psi}-\chi \chi)
\Big).
\end{multline}
The bulk equations of motion are
\begin{align} \label{bulkpsi}
0&= -i\overline{\sigma}\cdot \partial \psi - \partial_z \overline{\chi} +\frac{c}{z}\overline{\chi}+ \frac{\overline{d}}{z}\overline{\psi},\nonumber \\
0&=-i\sigma \cdot \partial \overline{\chi} - \partial_z \psi +\frac{c}{z}\psi- \frac{\overline{d}}{z}\chi.
\end{align}
In general, the 4-d mass eigenstates will correspond to Majorana fermions, $\xi_n (x)$, satisfying
\begin{gather}
-i\overline{\sigma}\cdot \partial \xi_n + m_n \overline{\xi_n}=0,
\end{gather}
such that the Kaluza-Klein expansion takes the form
\begin{align} \label{kkexp}
\psi &= \Sigma_n f_n (z)\xi_n (x), \nonumber \\
\chi &= \Sigma_n \overline{g}_n (z)\xi_n (x).
\end{align}
The bulk equations of motion then become
\begin{align}\label{bulkf}
0&= z{f'}_n +c f_n - \overline{d}\overline{g}_n-\overline{m}_n z g_n, \nonumber \\
0&=z{g'}_n -c g_n - \overline{d}\overline{f}_n+ m_n zf_n.
\end{align}
In the case of pure-Dirac bulk masses ($d=0$), these equations are easily decoupled; $f_n/\sqrt{z}$ and $g_n/\sqrt{z}$ 
are given by linear combinations of the two independent Bessel functions of order $c+1/2$ and $c-1/2$, respectively.
The case $d\neq 0$ is not so straightforward, as we illustrate by considering the special case where we take $d$ to be real.
Then, differentiating the second equation of (\ref{bulkf}) and substituting for $g_n$ and ${g'}_n$, we obtain the following uncoupled
equation for $f_n$,
\begin{gather} \label{fbulk}
{f''}_n + \left(\frac{1}{z}-\frac{1}{z-d/m_n}\right){f'}_n -\left(\frac{c^2+d^2}{z^2}+\frac{c-dm_n z + m_n^2 z^2}{z(z-d/m_n)}\right)f_n=0,
\end{gather}
with a similar equation for $g_n$. This has regular singular points at $z=0$ and $z=d/m_n$ in $z \; \epsilon \; \mathbb{C}$, and an
irregular singular point at $z=\infty$ (as one can show by substituting $z=1/\xi$). Thus, 
(\ref{fbulk}) is not Fuchsian, nor is it of confluent hypergeometric form, except when $d\rightarrow 0$ (in which case the two
regular singular points coalesce in a single regular singular point). 

Given that the origin is a regular singular point, we are able to find two independent series solutions about this point
by the method of Frobenius. These serve as suitable definitions for the independent solutions, at least for $|z|<d/m_n$. 
This will also turn out to be the most useful definition for phenomenological reasons, since the leading order behaviour of
the solutions near the orgin determines
the masses of ultralight modes, with $m_n z_1 \ll m_n z_0 \ll 1$.
Returning to the most general case where $d \; \epsilon \; \mathbb{C} $, we make the {\em ans\"{a}tze} 
\begin{gather} \label{series}
f_n = \Sigma_{m=0} a_{nm}^{(\alpha)}z^{\alpha +m},\;\; g_n =\Sigma_{m=0}b_{nm}^{(\beta)} z^{\beta +m},
\end{gather}
that yield
\begin{align}\label{frob}
0 &= \Sigma_{m=0} \left( (\alpha+ m + c)a_{nm}^{(\alpha)} z^{\alpha+m} +\overline{d} {\overline{b}}_{nm}^{(\beta)} z^{\overline{\beta}+m} -{\overline{m}}_nb_{nm}^{(\beta)} z^{\beta+ m+1}\right),\nonumber \\
0 &= \Sigma_{m=0} \left( (\beta+m -c)b_{nm}^{(\beta)} z^{\beta+m}+ \overline{d} {\overline{a}}_{nm}^{(\alpha)} z^{\overline{\alpha}+m} +m_n a_{nm}^{(\alpha)} z^{\alpha+ m+1}\right),
\end{align}
upon substitution.
The indicial equation, at $m=0$, has a solution with non-vanishing $a_{n0}^{(\alpha)}$ and $b_{n0}^{(\alpha)}$ iff.\ $\beta = \overline{\alpha}$, and
\begin{gather}
\alpha^2 = cc + d\overline{d} \;\; \Rightarrow \;\; b_{n0}^{(\beta)} = -\frac{\overline{\alpha}+c}{d}{\overline{a}}_{n0}^{(\alpha)}.
\end{gather}
The two roots $\alpha = \alpha_{\pm}$ are, real, equal in magnitude and opposite in sign. 
Let us re-define 
\begin{gather}
\alpha = +\sqrt{cc+d\overline{d}}=\overline{\alpha};
\end{gather}
Then, the two roots are given by $\alpha_{\pm}=\pm \alpha$.
The subsequent unknown coefficients $a_{nm>0}^{(\alpha)}$ and $b_{nm>0}^{(\alpha)}$ are determined iteratively by the analogue of (\ref{frob}),
but containing a linear superposition of the series for both roots $\alpha_{\pm}$, viz.
\begin{align}
0 &= \Sigma_{m=0} \Sigma_{\alpha = \alpha_{\pm}}\left( (\alpha+ m + c)a_{nm}^{(\alpha)} z^{\alpha+m} +\overline{d} {\overline{b}}_{nm}^{(\alpha)} z^{\overline{\beta}+m} -{\overline{m}}_nb_{nm}^{(\alpha)} z^{\beta+ m+1}\right),\nonumber \\
0 &= \Sigma_{m=0} \Sigma_{\alpha = \alpha_{\pm}}\left( (\beta+m -c)b_{nm}^{(\alpha)} z^{\beta+m} + \overline{d} {\overline{a}}_{nm}^{(\alpha)} z^{\overline{\alpha}+m} +m_n a_{nm}^{(\alpha)} z^{\alpha+ m+1}\right).
\end{align}
The coefficients of the two series with differing roots $\alpha_{\pm}$ are determined independently of each other. 
Thus we obtain two independent solutions for all $c$ and $d$ (with the exception of the 
special case where $\sqrt{c^2+d\overline{d}}$ is half-integral, and the second independent solution contains a logarithm).
We can use these series to define the independent solutions as far as the singularity at $|z| = |d/m_n|$, and presumably by analytic continuation beyond.

We can also solve the bulk equations of motion (\ref{bulkf}) at large values of $|m_n| z$. At leading order, they become
\begin{align}
{f'}_n & \sim   {\overline{m}}_n g_n,\nonumber \\
{g'}_n & \sim  - m_n f_n,
\end{align}
with solution
\begin{align}
f_n &\sim Az^p \sin |m_n|z + Bz^p \cos |m_n|z ,\nonumber \\
g_n &\sim -B z^p \sin |m_n|z + Az^p \cos |m_n|z.
\end{align}
At next-to-leading order, the equations (\ref{bulkf}) have consistent solution only if $p=0$, whence the asymptotic solutions at large argument are
\begin{align} \label{largearg}
f_n &\sim A \sin |m_n|z + B \cos |m_n|z ,\nonumber \\
g_n &\sim -B \sin |m_n|z + A \cos |m_n|z.
\end{align}
These can be matched onto the solutions at small argument, given a suitable analytic continuation.
\subsection{\label{bc}Boundary Dynamics}
We wish to consider the boundary conditions at $z=z_0$ and $z=z_1$ that result from dynamics localized on the UV and IR branes.
In particular, we shall be interested in the boundary conditions at $z=z_0$ that result from the UV-brane-localized interactions between the fermions in the sterile throat
and fields in the SM throat. This is most conveniently done following the formalism of \cite{Csaki:2003sh}. 
To wit, we consider the theory compactified on the interval $z \; \epsilon \; [z_0,z_1] $, as opposed to an orbifold.
The bulk equations of motion (\ref{bulkpsi}) are obtained from the vanishing of the variation of the action (\ref{actpsi}), which also includes the boundary variation 
\begin{gather}
\delta S =  \frac{1}{2}\int d^4 x \; \Big[
\delta\chi \psi - \delta\overline{\psi} \overline{\chi} +\delta\overline{\chi}\overline{\psi}-\delta \psi \chi
\Big]_{z_0}^{z_1}.
\end{gather}
This boundary variation vanishes iff.\ 
\begin{gather}\label{genbc}
\chi_{\alpha} = M_{\alpha}^{\beta}\psi_{\beta} + N_{\alpha \dot{\beta}}{\overline{\psi}}^{\dot{\beta}},
\end{gather}
at $z=z_0,z_1$, where
\begin{align}
M_{\alpha}^{\beta} &=\pm \epsilon^{\beta\gamma}M_{\gamma}^{\delta}\epsilon_{\delta\alpha},\nonumber \\
N_{\alpha}^{\dot{\beta}} &= \pm {N^{\dagger}}_{\alpha}^{\dot{\beta}}.
\end{align}
($M_{\alpha}^{\beta}$ and $N_{\alpha}^{\dot{\beta}}$ may include boundary derivatives, and the $\pm$ signs account for partial integrations of these.)
What are the particular forms of the general boundary condition (\ref{genbc}) that result from adding brane-localized terms to the action? At least in simple cases, this question can be 
answered unambiguously  \cite{Csaki:2003sh} by first adding the localized terms in the bulk at a distance $\epsilon$ away from the relevant brane, {\em viz.\ }at $z=z_0^+\equiv z_0 + \epsilon$ or at $z=z_1^-\equiv z_1 -\epsilon$ and solving 
the jump equations there. Then, one matches the solution to the original boundary conditions at $z=z_0$ or $z_1$ and takes the limit $\epsilon \rightarrow 0$.
This slightly convoluted procedure avoids dealing with functions that are discontinuous at the boundary.

As an example, let us consider how the standard boundary condition $\chi |_{z_0}=0$ is modified by a Dirac mass term mixing the field $\psi$
and a brane-localized fermion $\eta(x)$.\footnote{In the holographic dual \cite{Contino:2004vy}, $\psi(x,z_0)$ represents a dynamical source field, external to the CFT, that mixes with the CFT and with the external field $\eta(x)$.}
We add to the action (\ref{actpsi}) the UV-localized term
\begin{gather} \label{uvact}
\delta S = - \int d^4 x \; \left( -i \eta \sigma \cdot \partial \overline{\eta} + bk^{1/2}\eta \psi + \overline{b}k^{1/2}\overline{\psi}\overline{\eta} \right)\Big|_{z_0^+},
\end{gather}
which is manifestly Hermitian and 4-d Lorentz-invariant.
Later on, we will take $\eta$ to be some linear combination of the SM neutrinos.
The new bulk equations of motion are
\begin{align}\label{bulkpsimod}
0&= -i\overline{\sigma}\cdot \partial \psi - \partial_z \overline{\chi} +\frac{c}{z}\overline{\chi}+ \frac{\overline{d}}{z}\overline{\psi}+\delta (z-z_0^+)\overline{b}k^{1/2}\overline{\eta},\nonumber \\
0&=-i\sigma \cdot \partial \overline{\chi} - \partial_z \psi +\frac{c}{z}\psi- \frac{\overline{d}}{z}\chi, \nonumber \\
0&=\delta (z-z_0^+)(-i \sigma \cdot \partial \overline{\eta} + bk^{1/2}\psi ).
\end{align}
Assuming that $\eta$ and $\psi$ are continuous at $z_0^+$, the jump equations are
\begin{align}
0 &= -[\overline{\chi}]_{z_0^+} +\overline{b}k^{1/2}\overline{\eta}\big|_{z^+_0}, \nonumber \\
0 &= -i \sigma \cdot \partial [\overline{\chi}]_{z_0^+} + [\partial_z \psi]_{z_0^+},
\end{align}
where $[\overline{\chi}]_{z_0^+}$ denotes the jump in $\overline{\chi}$ at $z=z_0^+$. Matching to the original boundary condition $\chi |_{z_0}=0$,
and taking the limit $\epsilon \rightarrow 0$, these reduce to the second bulk equation of motion (\ref{bulkpsi}) evaluated at the boundary, together with 
the new boundary condition 
\begin{gather} \label{diracbcpsi}
0 = \left(-\overline{\chi}+\overline{b}k^{1/2}\overline{\eta}\right)\Big|_{z_0}.
\end{gather}
We may use the third equation of (\ref{bulkpsimod}) to write this condition purely in terms of bulk fields as
\begin{gather}\label{diracbc}
0= \left(-i \sigma \cdot \partial \overline{\chi} + \overline{b}bk\psi \right)\Big|_{z_0}.
\end{gather}
Since the bulk fermions are gauge singlets, we are also free to add Majorana mass terms, 
localized on either the UV or IR branes, of the form
\begin{gather}
\delta S = - \int d^4 x \; \left(a \psi \psi  + \mathrm{H.\ c.}\right)\Big|_{z_0^+,z_1^-},
\end{gather}
which leads to a modified boundary condition of the form
\begin{gather}
\delta S = \left(-\chi + a \psi \right)\Big|_{z_0^+,z_1^-}.
\end{gather}
Such terms present no added difficulty in terms of computation, but nor do they change the character of the results. 
We disregard them in what follows for the sake of clarity.
\subsection{The Ultralight Mode}
In the case of pure-Dirac bulk masses, $d=0$, the boundary conditions $\chi |_{z=z_0,z_1}=0$ (or $\psi |_{z=z_0,z_1}=0$)
result in a massless zero mode. No such massless mode is obtained with $d \neq 0$, as we now show. For $m_0=0$, the general solution to the bulk equations of motion (\ref{bulkf})
is given exactly by the leading order part of the series solution (\ref{series}), which we write as
\begin{align} \label{smallarg}
f_0 &= a_+ z^{\alpha} + a_- z^{-\alpha}, \nonumber \\
g_0 &= -\frac{\overline{\alpha}+c}{d}{\overline{a}}_+ z^{\overline{\alpha}} 
 -\frac{-\overline{\alpha}+c}{d}{\overline{a}}_- z^{-\overline{\alpha}}.
\end{align}
A non-trivial solution with one of $a_{\pm}$ non-vanishing, and satisfying the boundary condition $\chi |_{z=z_0,z_1} \Rightarrow g_0 |_{z=z_0,z_1} =0$
is possible only if $\overline\alpha = \pm c \Rightarrow d=0$, implying that zero modes are only obtained in the pure-Dirac limit.
The same argument shows that no modes with masses $m_n z_1 \ll 1$ are present given such boundary conditions, and we conclude that the
spectrum begins with modes whose mass is at least of order $1/z_1$.

What happens if we add a UV-brane-localized fermion $\eta$? The boundary condition at $z=z_0$ is changed to (\ref{diracbc}), or equivalently
\begin{gather} \label{diracbcf}
0 = -{\overline{m}}_n g_n(z_0) + \overline{b}bk f_n(z_0),
\end{gather}
together with the original boundary condition at $z=z_1$, namely
\begin{gather} \label{neubcf}
0 =  g_n(z_1).
\end{gather}
Now, as $b\rightarrow 0$, (\ref{diracbcf}) is trivially solved by $m_0=0$, so that we are guaranteed the existence of a massless mode. 
This is no surprise: as $b\rightarrow 0$, the fermion $\eta$ becomes decoupled from the bulk fermion $\psi$ and is manifestly massless.
Similarly, for sufficiently small values of $b$, we expect to find an ultralight mode in the spectrum with $m_0 z_1 \ll 1$.
To find the mass of this mode, we can use the small argument solutions of (\ref{smallarg}), in terms of which the 
modified boundary conditions become
\begin{align}
0 &= \frac{\overline{m}_0}{d}((\overline{\alpha}+c){\overline{a}}_+ z_0^{\overline{\alpha}}+
(-\overline{\alpha}+c){\overline{a}}_- z_0^{-\overline{\alpha}})+\overline{b}bk(a_+ z_0^{\alpha} + a_- z_0^{-\alpha}),\nonumber \\
0 &= -\frac{\overline{\alpha}+c}{d}{\overline{a}}_+ z_1^{\overline{\alpha}} 
-\frac{-\overline{\alpha}+c}{d}{\overline{a}}_- z_1^{-\overline{\alpha}}.
\end{align}
After some algebra, we find that these equations have non-trivial solutions iff.\ $m_0$ is such that
\begin{gather}
|m_0 |^2 = |b|^4 k^2\frac{c+\alpha}{c-\alpha}.
\end{gather}
This expression is manifestly real and positive.
We find in either case that the mass is of order $|b|^2k$, so is ultralight only if $b^2 \ll z_0/z_1$. We shall learn in the sequel that $b$ naturally
takes values of order $\mathrm{TeV}/k$, so we get an ultralight mode if $z_1^{-1}\gg (\mathrm{TeV})^2/k$. Note also that the relative fractions of 
the bulk and brane-localized fields that make up the ultralight mode are $\psi : \chi : \eta \sim b : b : 1$, so that the ultralight mode is mostly active
and the higher KK modes will be mostly sterile.

Can we say anything more about the spectrum of higher KK modes? For low values of $n$, we would need to solve the bulk equations of motion
(\ref{bulkf}) with given boundary conditions numerically to find the spectrum. However, at larger KK numbers, such that
$z_1^{-1}\ll m_n \ll z_0^{-1}$, we can approximate the solutions to (\ref{bulkf}) in the region of the IR brane by the large argument
solutions (\ref{largearg}), namely
\begin{align}
f_n &\sim A \sin |m_n|z + B \cos |m_n|z ,\nonumber \\
g_n &\sim -B \sin |m_n|z + A \cos |m_n|z.
\end{align}
Now, a general boundary condition on the IR brane of the form $(\chi + a \psi)|_{z_1}=0$, where $a$ is a constant,
will lead to an expression for the ratio of the unknowns $A/B$ as a rational linear function of $\tan |m_n|z_1$. 
The allowed values of $|m_n|$ are those for which this ratio coincides with the ratio of the unknowns on the UV brane, which will be given by some function of
$|m_n|z_0$, $f(|m_n|z_0)$ say. Imagine that $|m_n|$ is such a value. If $|m_n|$ is increased by $\pi/z_1$, the rational function
returns to the same value, whereas $f(|m_n|z_0)$ changes by a negligible amount: $f(|m_n|z_0) \rightarrow f(|m_n|z_0)+ \frac{\pi z_0}{z_1}f'(|m_n|z_0)$. 
Thus, up to corrections of $O(z_0/z_1)$, we see that KK modes at large $n$ are separated in mass by an amount $\pi/z_1$.

\section{\label{>1BN}Multiple Bulk Sterile Neutrinos}
With just one bulk sterile neutrino, we obtain only one massive, but ultralight, mode, which is mostly composed 
of a linear combination of the active neutrinos. However, the solar and atmospheric neutrino data indicate two independent mass-squared
differences, so a realistic sterile throat model should contain at least two bulk fermions in the sterile throat.\footnote{ Or, alternatively, more than one sterile throat.}
With this in mind, we now generalize the formalism to the case of multiple bulk 
sterile neutrinos $\Psi^i$ coupled to multiple brane-localized fermions $\eta^I$, the  bulk action generalizes to
\begin{multline} \label{actpsi2}
S = - \int d^4 x dz \; \Bigg( 
-i\chi^i \sigma \cdot \partial \overline{\chi}^i
-i\overline{\psi}^i\overline{\sigma}\cdot \partial \psi^i
+\frac{1}{2}\left( \chi^i \partial_z \psi^i + \partial_z \overline{\psi}^i \overline{\chi}^i - \overline{\psi}^i\partial_z \overline{\chi}^i-\partial_z \chi^i \psi^i \right)\nonumber \\
+\frac{c_{ij}}{z}(\chi^i \psi^j + \overline{\psi}^i \overline{\chi}^j) 
+\frac{d_{ij}}{2z}(\psi^i \psi^j - \overline{\chi}^i \overline{\chi}^j) +\frac{\overline{d}_{ij}}{2z}(\overline{\psi}^i\overline{\psi}^j-\chi^i \chi^j)
\Bigg),
\end{multline}
where the matrices $c_{ij}$ and $d_{ij}$ are Hermitian and symmetric, respectively. To this we add the boundary action
\begin{gather}
\delta S = - \int d^4 x \; \left( -i \eta^I \sigma \cdot \partial \overline{\eta}^I + b_{Ii}k^{1/2}\eta^I \psi^i + b^{\dagger}_{iI}\overline{\psi}^i\overline{\eta}^I \right)\Big|_{z_0^+}.
\end{gather}
where $b_{Ii}$ is a complex matrix. The Hermitian matrix $c_{ij}$ can be diagonalized by a unitary transformation that acts identically on $\psi^i$ and $\chi^i$,
but the symmetric matrix $d_{ij}$ will not in general be diagonal in this basis. Similarly, if we choose a basis in which the masses of
charged leptons are diagonal, then we must also consider a general form for the matrix $b_{Ii}$. 

As in the case of one bulk singlet fermion, we solve for the mass eigenstates by expanding
\begin{align}
\psi^i &= \Sigma_n f^i_n (z)\xi_n (x), \nonumber \\
\chi^i &= \Sigma_n \overline{g}^i_n (z)\xi_n (x).
\end{align}
A series solution for the bulk equations of motion generalizing (\ref{bulkf}) is then given by
\begin{align}\label{ser}
f_n^i &= \Sigma_{m,\lambda} a_{nm}^{({\lambda})i}z^{\alpha_{\lambda} +m}, \nonumber \\
g_n^i &= \Sigma_{m,\lambda} b_{nm}^{({\lambda})i}z^{\overline{\alpha_{\lambda}} +m},
\end{align}
where $\alpha_{\lambda}$ are the eigenvalues of the matrix
\begin{gather}
\begin{pmatrix}
c^i \delta^{ij} & d^{ij}\\  \overline{d}^{ij} & -c^i \delta^{ij}
\end{pmatrix}.
\end{gather}
The boundary conditions are, at $z=z_1$,
\begin{gather}
\chi^i |_{z_1} = 0,
\end{gather}
and, at $z=z_0$,
\begin{gather}
\left(-i \sigma \cdot \partial \overline{\chi}^i + \frac{(b^{\dagger}b)_{ij}}{z_0}\psi^j\right)\Big|_{z_0} = 0.
\end{gather}
Substituting from (\ref{ser}), the determination of the masses of the ultralight modes is simply 
an exercise in linear algebra, albeit somewhat involved in the general case. To illustrate the character of the result,
we find the masses in the simpler case where the mass matrices $c$ and $d$ are simultaneously diagonalizable.
Then the bulk equations of motion for different $i$ decouple and the solutions at small arguments, satisfying the IR boundary condition, are approximately
\begin{align}
f_n^i &= a^i \overline{d}^i \left(-\frac{1}{c^i + \alpha^i}(\frac{z}{z_1})^{\alpha^i}+ \frac{1}{c^i - \alpha^i}(\frac{z}{z_1})^{-\alpha^i}
\right), \nonumber \\
g_n^i &= \overline{a}^i  \left((\frac{z}{z_1})^{\overline{\alpha}^i}
-(\frac{z}{z_1})^{-\overline{\alpha}^i}\right).
\end{align}
The UV boundary condition then reads
\begin{gather}
\overline{m}_n  \overline{a}^i = B_{ij} a^j,
\end{gather}
where
\begin{gather}
B_{ij} = \frac{(b^{\dagger}b)_{ij}}{d^j z_0}\frac{((c^j + \alpha^j)(\frac{z_0}{z_1})^{-\alpha^j}-(c^j - \alpha^j)(\frac{z_0}{z_1})^{\alpha^j})}{(\frac{z_0}{z_1})^{\overline{\alpha}^i}
-(\frac{z_0}{z_1})^{-\overline{\alpha}^i}}.
\end{gather}
Now, since 
\begin{gather}
\overline{m}_n \overline{a}^i = B_{ij} a^j \Leftrightarrow m_n a^i = \overline{B}_{ij} \overline{a}^j
\end{gather}
we find that $|m_n|^2$ is an eigenvalue of the matrix $\overline{B}B$. By considering the $b \rightarrow 0$ limit, we see that there will be
one ultralight mode for each brane fermion, each with mass of $O(b^2/z_0)$, where $b$ is a typical element 
of the matrix $b_{Ii}$. The ultralight modes are predominantly composed of active neutrinos, with O(1) mixing between the flavour eigenstates.
\subsection{Oscillation Phenomena}
The pattern of neutrino masses and mixings that arises for two or more bulk sterile neutrinos
is qualitatively similar to that which is observed in so-called $3+2$ models.
These have three ultralight modes, which are mostly-active, with $O(1)$ mixings
between the three active neutrino flavours of the SM. On top of this, there are two additional, light neutrino states, which are mostly sterile.

In the $3+2$ models, it is argued that this structure gives a good fit to the oscillation data, with solar and atmospheric data
explained by oscillations between the ultralight modes, and the LSND result explained by oscillations occurring via the mostly-sterile states.

The sterile throat models are similar, with the major difference that there are
now a huge number of mostly-sterile states (the higher KK modes).
The beam energy at LSND is roughly an MeV, meaning that around $10^6$ mass eigenstates in the KK tower are available for oscillation.
Because the KK masses are not quite commensurate,
the oscillation signal will appear to be essentially aperiodic \cite{Dienes:1998sb}, but 
this is of little relevance for LSND, which does not resolve the oscillations.
Indeed, roughly-speaking, the only criterion that LSND imposes on the mass scale (or scales) is that it be large enough such that 
significant oscillations occur over the short baseline. This is guaranteed by choosing a large-enough KK scale.

Can a sterile throat model explain the spectrum of solar and atmospheric mass-squared differences?
We found above that the ultralight modes have masses of order $b^2/z_0$. The coefficient $b$ comes from
the Dirac mass term localized on the UV brane and its size, assuming Yukawa couplings of order unity, is set by the expectation
value of the Higgs field on the UV brane. 

Now, in the most realistic models of warped electroweak symmetry breaking \cite{Contino:2003ve,Agashe:2004rs},
the 4d Higgs field corresponds to the zero mode of the fifth component of a bulk gauge field. 
This set-up is holographically dual to realizing the Higgs as a pseudo-Goldstone boson in 4d, and protects the Higgs
mass parameter from dangerous radiative corrections down to the KK scale.
The zero mode has an IR-localized profile $A_5(y,x) \simeq y A_5 (x)$, where $y$ is the fifth co-ordinate in the SM throat. 
At the UV brane, the Higgs vev is therefore expontentially suppressed relative to its vev on the SM IR brane (which is $k$ in 5d units)
and $b \sim$TeV.\footnote{In models where the Higgs is the light mode of a bulk scalar \cite{Davoudiasl:2005uu},
the extra-dimensional profile of the bulk Higgs, and ergo the {\em vev} on the UV brane, can be tuned. This gives further freedom in the value of $b$.}

Alternatively, if the SM is simply taken to be localized on the UV brane, we see directly that $b$ has
to be of order TeV, so as to get the correct masses for SM fields. This value is, of course, unnatural though.

Thus to get the right masses for the ultralight modes, we need to choose $k = 1/z_0 \sim 10^{14-15}$GeV. 
Normally, when one considers warped compactifications, one simply assumes that all fundamental 5d mass scales are
roughly the same. If we were to apply this logic here, we would conclude that the 4d Planck scale, $M_4$ would be too low.
However, we argue that not only is a small hierarchy between $k$ and $M_4$ desirable here, but also that it is 
essential for 
the consistency of the 5d effective field theory formalism that we have employed throughout.

Indeed, the 5d effective field theory is valid up to the scale at which 5d gravity becomes strongly coupled, $M_5$. In order
to be able to trust the warped geometry solutions of the 5d Einstein field equations, we require that the bulk
cosmological constant and brane tensions, $\Lambda^{1/5}_5$ and $\pm \Lambda^{1/4}_4$, should be much less than $M_5$.
Since the curvature scale $k$ and the 4d Planck scale are given by
\begin{align}
k^2 &= M_5^2 \left(\frac{\Lambda^{1/5}_5}{M_5}\right)^5,\nonumber \\
M_4^2 &= M_5^2 \left(\frac{\Lambda^{1/5}_5}{M_5}\right)^{-5/2},
\end{align}
respectively, we see that a hierarchy between the two is obligatory, and that $\Lambda^{1/5}_5/M_5 \sim 10^{-1-2}$ or so suffices to reproduce the correct 
Planck scale.\footnote{Note that the cut-offs of the 5d theory ($M_5$) and its 4d dual ($k$) need not be and are not the same.
The correspondence is only valid up $k$, because at shorter scales the extra dimensions of the $AdS_5 \times S_5$ (or similar)
manifold show up, and the simple 5d/4d correspondence breaks down. Since the strongly-coupled CFT is defined by its weakly-coupled dual,
the cut-off of the 4d theory is necessarily $k$. In the multi-throat case, the correspondence breaks down at the smallest curvature scale.}
We thus conclude that masses for the ultralight modes in the range consistent with solar and atmospheric oscillation data can be easily achieved.
These oscillations then occur via the order-one mixing between the ultralight modes.
\section{\label{highp}High-Energy Behaviour}
We have seen above that sterile neutrinos in a separate throat to the SM may provide an explanation for all neutrino oscillation data,
with natural explanations for light and ultralight masses in terms of a high scale, via the mechanisms of warping and a higher-dimensional see-saw.
We now consider the effect of the extra KK towers of mostly-sterile neutrinos on physics at high energies,
in particular in the early Universe.

As far as CFT interactions are concerned, a key observation is that, as we show below, the extra states (the higher KK modes) are all localized in the IR.
This means that interactions with SM fields (which occur only on the UV brane) can proceed only via the 5d UV-to-IR-brane
propagator for some bulk field that interacts with the mostly-sterile higher KK fermion modes.
Examples of such fields include the bulk sterile neutrinos themselves, the graviton, or some other SM-gauge-singlet field that propagates in the sterile throat.
A generic property of such propagators in a slice of AdS is that they decay exponentially for values of the
Euclidean four-momentum far above the IR scale $1/z_1$. 
We will show this explicitly for bulk sterile neutrinos with non-zero bulk Majorana masses, subject to the boundary conditions considered previously.
It has already been shown for other bulk fields in \cite{Gherghetta:2000kr}.
This means that equilibriation processes involving interactions in the CFT are negligible at energies and momenta
much above an eV. In the 4-d dual, the explanation for this fall-off is that such interactions proceed through
the strong dynamics of the CFT, which involves a form factor.
The CFT in transparent to probes on scales much shorter than an inverse-eV, which accounts for the fall-off of the form factor.

This conclusion does not hold for SM interaction processes, in which equilibrium is attained via pure SM interactions and transmitted
to the sterile states via oscillations. These oscillations occur via the two-point `interactions' corresponding
to the couplings between the SM neutrinos and the elementary source field (dual to $\psi (z_0,x)$),
and between the source and the CFT. These two-point couplings do not go through the strong dynamics of the CFT and do not
come with a form factor. 
\subsection{Localization Properties of the Modes}
To show that the higher KK modes are localized in the IR, we show that they are normalizable even as the UV brane is removed, 
that is, in the limit as $z_0 \rightarrow 0$. We consider the case of a single bulk sterile neutrino for simplicity.
In order to canonically normalize the kinetic terms for $\xi_n (x)$, we see from (\ref{actpsi}), (\ref{kkexp}) and (\ref{uvact}) that
\begin{gather} \label{norm}
1 = \frac{1}{\overline{b}b k}|g_n(z_0)|^2 + \int_{z_0}^{z_1} dz \; \left( |f_n(z)|^2 + |g_n(z)|^2 \right),
\end{gather}
where $f_n(z)$ and $g_n(z)$ are solutions of (\ref{bulkf}) satisfying the boundary conditions (\ref{diracbcf}) and $g_n(z_1)=0$.
For modes such that $m_n$ remains finite as $z_0 \rightarrow 0$ (namely all modes with finite $n$ bar the ultralight mode with $n=0$),
the boundary conditions reduce to $g_n (z_1) = 0$ and $f_n(z_0)=0$, and the normalization integral (\ref{norm}) reduces to 
\begin{gather} \label{norm2}
\int_{z_0}^{z_1} dz \; \left( |f_n(z)|^2 + |g_n(z)|^2 \right).
\end{gather}
Near $z=z_0$, we can approximate the solutions $f_n$ and $g_n$ by their values at small arguments. They thus take the form
\begin{align}
f_n &= A \left( (\frac{z}{z_0})^{\alpha} + (\frac{z}{z_0})^{-\alpha}\right),\nonumber \\
g_n &= - A \left( \frac{\alpha+c}{\overline{d}}(\frac{z}{z_0})^{\alpha} + \frac{\alpha-c}{\overline{d}}(\frac{z}{z_0})^{-\alpha}\right).
\end{align}
Substituting these solutions into (\ref{norm2}), it is clear that the integral is finite as $z_0 \rightarrow 0$. 
Therefore all modes with $n \neq 0$ are normalizable as $z_0 \rightarrow 0$ and are localized in the IR.
We repeat that this conclusion does not hold for the 
ultralight mode with $n=0$, for which $m_0 z_0$ is finite as $z_0 \rightarrow 0$.\footnote{The ultralight mode is, in general, localizable nowhere.
This is similar to the zero mode of a bulk gauge field, or a Dirac fermion with $c=1/2$.}
\subsection{The Brane-to-Brane Propagator}
We now calculate the UV-to-IR brane 5d-propagator for the bulk sterile neutrino at large 4-momenta, showing that it is exponentially suppressed for
Euclidean momenta above the IR scale $1/z_1$. At large 4-momenta, the bulk equations of motion (\ref{bulkpsi}) 
can be self-consistently truncated as
\begin{align}
0 &= i \sigma \cdot \partial \psi + \partial_z \overline{\chi}, \nonumber \\
0 &= - i \overline{\sigma}\cdot \overline{\chi} + \partial_z \psi.
\end{align}
We therefore define the Green's functions at large Minkowskian 4-momenta as
\begin{gather}\label{green}
\begin{pmatrix}
\delta^{\dot{\alpha}}_{\dot{\beta}} \partial_z & -\overline{\sigma}^{\mu\dot{\alpha}\beta} p_{\mu}\\ \sigma_{\alpha\dot{\beta}}^{\mu} p_{\mu}& \delta_{\alpha}^{\beta} \partial_z
\end{pmatrix} 
\begin{pmatrix}
S_{\dot{\gamma}}^{\dot{\beta}} & S^{\dot{\beta}\gamma} \\ S_{\beta \dot{\gamma}} & S_{\beta}^{\gamma}
\end{pmatrix}(z,z',p )=
\begin{pmatrix}
\delta^{\dot{\alpha}}_{\dot{\gamma}} & 0 \\ 0 & \delta^{\gamma}_{\alpha},
\end{pmatrix} \delta (z-z'),
\end{gather}
subject to the boundary conditions
\begin{align}
S_{\dot{\gamma}}^{\dot{\beta}} (z_1,z',p) &= S^{\dot{\beta}\gamma}(z_1,z',p) = 0, \nonumber \\
\sigma_{\alpha\dot{\beta}}^{\mu} p_{\mu}  S_{\dot{\gamma}}^{\dot{\beta}} (z_0,z',p) +\frac{\overline{b}b}{z_0} S_{\alpha \dot{\gamma}} (z_0,z',p) &= 
\sigma_{\alpha\dot{\beta}}^{\mu} p_{\mu}  S^{\dot{\beta}\gamma} (z_0,z',p) +\frac{\overline{b}b}{z_0} S_{\alpha}^{\gamma} (z_0,z',p)= 0 .
\end{align}
The Green's functions defined in (\ref{green}) are given by
\begin{gather}
\begin{pmatrix}
S_{\dot{\gamma}}^{\dot{\beta}} & S^{\dot{\beta}\gamma} \\ S_{\beta \dot{\gamma}} & S_{\beta}^{\gamma} 
\end{pmatrix}=
\begin{pmatrix}
\delta_{\dot{\gamma}}^{\dot{\beta}} \partial_z G_1 & \overline{\sigma}^{\nu \dot{\beta}\gamma}p_{\nu}G_2 \\ -\sigma^{\nu}_{\beta \dot{\gamma}}p_{\nu}G_1 & \delta_{\beta}^{\gamma} \partial_z G_2
\end{pmatrix},
\end{gather}
where $G_{1,2}$ are themselves Green's function solutions of
\begin{gather}
(\partial_z^2 - p^2)G_{1,2}(z,z',p) = \delta (z-z').
\end{gather}
After some tedious algebra (which we spare the reader), we obtain the UV-to-IR brane propagators for spacelike Minkowskian momenta 
\begin{gather}
\begin{pmatrix}
S_{\dot{\gamma}}^{\dot{\beta}} & S^{\dot{\beta}\gamma} \\ S_{\beta \dot{\gamma}} & S_{\beta}^{\gamma} 
\end{pmatrix}(z_0,z_1,p)=
\begin{pmatrix}
\delta_{\dot{\gamma}}^{\dot{\beta}} \frac{1-C}{-1+Ce^{-2pz_1}}e^{-pz_1} & 0 \\ -\sigma^{\nu}_{\beta \dot{\gamma}}\frac{p_{\nu}}{p} \frac{1+C}{-1+Ce^{-2pz_1}}e^{-pz_1}& 0
\end{pmatrix},
\end{gather}
where $C=\frac{p z_0 - \overline{b}b}{p z_0 +\overline{b}b}$. Performing the analytic continuation $p_0 \rightarrow i p_0$, we see that 
the propagators do indeed fall off exponentially for Euclidean 4-momenta greater that the IR scale $1/z_1$.
\subsection{High-Energy Phenomenology}
The exponential fall-off of the bulk propagators essentially means that CFT interaction processes
involving the higher KK states and SM fields have negligible cross-section at energies much above an eV. Indeed, one can imagine that there are enhancement factors, such as mass threshold 
and phase space effects, which grow as powers of the energy, but these will always be overwhelmed by the exponential suppression of the propagator at large energy.

In particular, we see that the extra states can never come into thermal equilibrium with SM fields in the early Universe
through processes of this kind, at least when the temperature of SM fields is above an eV. By the time the temperature reaches an eV and the extra states have
a chance of reaching equilibrium, nucleosynthesis will be complete, and structure formation will already have begun. 

The extra states could however, come into equilibrium via SM interaction processes.
In order to assess to what extent the extra KK states reach equilibrium in this way, one would need  
to perform a numerical solution of the Boltzmann equations in the early Universe, taking into account the effects of neutrino mixing
and the multiple neutrino mass thresholds.

We make just one remark in this regard, which is that even if there are multiple massive states in complete equilibrium
with SM fields, they are not necessarily in contradiction with WMAP data. This is because increases in either $\Sigma_{\nu}m_{\nu}$ or $N_{\nu}$
can be compensated in the data fit, by moving the present Hubble parameter in opposite directions. Thus, increasing
both simultaneously has a compensatory effect, reflected for example in (\ref{summ}) \cite{Hannestad:2003xv}.
\section{Discussion}
In the foregoing, we have examined the impact of SM-gauge-singlet fermions (a.k.a. sterile neutrinos)
propagating in the bulk of a separate throat on physics in the SM throat. We find
that the bulk sterile neutrinos give rise to three, mostly-active, ultralight Majorana neutrinos in 4-d, with mass of 
order $\mathrm{TeV}^2 z_0$, where $1/z_0$ is the scale of the UV brane, or equivalently the cut-off of the dual CFT.
The bulk sterile neutrinos also lead to KK towers of mostly-sterile Majorana modes, beginning at the IR brane scale, or
equivalently the IR strong-coupling scale of the CFT.
There is order-one mixing in the mass eigenstates between the SM flavour eigenstates.
What is more, these conclusions hold for all order-one values of the dimensionless mass parameters in the bulk or on the boundary:
the phenomenology is quite generic. 

We have suggested that such a set-up could plausibly explain the entirety of the existing neutrino data, including LSND and cosmological
observations. We remark in passing that there is another experiment, the Heidelberg-Moscow experiment, 
which in one analysis \cite{Klapdor-Kleingrothaus:2001ke}
claims a signal for neutrinoless double-beta decay, interpreted as evidence for Majorana neutrino masses. 
The measured effective neutrino mass $|\Sigma_{\nu} m_{\nu} U^2_{e\nu}|$, where $U_{e\nu}$ is the mixing matrix element 
between the electron-neutrino and the mass eigenstates, is given by the ranges 0.11 -- 0.56 eV \cite{Klapdor-Kleingrothaus:2001ke}
or 0.4 -- 1.3 eV \cite{Yao:2006px}, depending on the values one takes for the relevant nuclear matrix elements.
These values could also be explained by a sterile throat scenario, with multiple states of eV mass and small matrix elements.

The next step in the analysis of sterile-throat models is a detailed numerical fit to the data. Is this feasible?
As regards LSND, or physics in the early Universe at MeV energies, this in principle involves finding the masses and mixing matrix elements
of around $10^6$ KK modes, given a spectrum beginning at around an eV.
However, the active component of the KK mass eigenstates goes to zero at large KK number, so we expect that including only 
a relatively small number of modes will prove to be a good approximation in practice.

Supposing that a sterile throat model does fit the existing data, how might we find conclusive evidence for it?
The oustanding signature of an extra-dimensional explanation of LSND is the KK tower beginning, at low energy. This has a dramatic effect
on the periodicity of oscillations, which appear essentially aperiodic \cite{Dienes:1998sb}. Whether any neutrino beam experiment 
will ever be able to resolve the oscillations sufficiently to observe this seems rather doubtful, but the observation of
even a few oscillation components with mass-squared differences of similar order would already be a strong hint.
Even if the presence of a KK tower in the neutrino oscillation data could be inferred, we would still be faced with the question of the nature of 
the relevant extra dimension: would it be warped, or flat, for example? These two scenarios are easily distinguished if the 
the masses of the first few KK states can be measured with reasonable precision, as they show integer 
spacing in the flat case \cite{Arkani-Hamed:1998vp,Dienes:1998sb}, but do not in the warped case.

Another possible effect of the KK tower is on high-energy astrophysical processes, for example invisible energy loss in stars
and supernovae. Such objects have temperatures corresponding to a large number of KK states being above threshold,
albeit with tiny individual couplings. It would be very interesting to explore the collective effects on such processes
of a sterile throat model. Again though, detailed numerical simulations are required. We hope to address this in future work.

Finally, we remark that the idea of an additional sterile throat, but with other bulk fields, may lead to other
interesting phenomenology. As examples, we suggest the axion \cite{Flacke:2006ad} or quintessence fields.

\begin{acknowledgments}
I thank T.~Flacke, J.~March-Russell, D.~Maybury, G.~G.~Ross and S.~M.~West for useful discussions.
\end{acknowledgments}

\end{document}